\documentclass[letterpaper, 10 pt, conference]{ieeeconf}  % Comment this line out
\IEEEoverridecommandlockouts                              % This command is only
\usepackage{graphicx}
\usepackage{float}
\usepackage{subfig}
\usepackage{url}
\usepackage{amsmath}
\newenvironment{mat}{\left[ \begin{array}{ccccccccccccc}}{\end{array}\right]}
\newcommand\bcm{\begin{mat}}
\newcommand\ecm{\end{mat}}
\usepackage{tikz}
\usetikzlibrary{graphs,graphs.standard}
\title{\LARGE \bf
How does a coin toss ?
}

\author{Jithin D. George
}

\begin{document}

\maketitle
\thispagestyle{empty}
\pagestyle{empty}

%%%%%%%%%%%%%%%%%%%%%%%%%%%%%%%%%%%%%%%%%%%%%%%%%%%%%%%%%%%%%%%%%%%%%%%%%%%%%%%%
\begin{abstract}

Is flipping a coin a deterministic process or a random one? We do not allow bounces. If we know the initial velocity and the spin given to the coin, mechanics should predict the face it lands on. However, the coin toss has been everyone's introduction to probability and has been assumed to be the hallmark random process. So, what's going on here?

\end{abstract}

%%%%%%%%%%%%%%%%%%%%%%%%%%%%%%%%%%%%%%%%%%%%%%%%%%%%%%%%%%%%%%%%%%%%%%%%%%%%%%%%
\section{Preface}
This article is an exploration of the problem described by Keller in [1]. Keller's idea serves as inspiration but restating his proof does not reveal anything different from [1]. So, here, a tangential perspective is explored starting from section 4. All the figures are an implementation of the theory described here and can be explored through the IPython Notebook [2].
\section{Introduction}

If we know the initial velocity of the coin and its initial angular momentum, it should be a deterministic problem to find the face it lands on. Let y(t) be the height of the coin at time $t$, $u$ and $w$ be the initial velocity and the angular velocity imparted to the coin and $g$ be the acceleration due to gravity.
We can find the height of the coin and the orientation of the coin at any time by these equations.
\[y(t) = ut - \frac{1}{2} gt^2 +y(0) \]
\[\theta(t) = \omega t\]
For the rest of this article, we will assume that we start with heads up and returns to the height it started from without any bounces. If the time of landing is $t_0$, we end up with heads up if 
 \[ 2n\pi - \frac{\pi}{2}<\theta(t_0) <  2n\pi + \frac{\pi}{2}\]
 
 That is,
  \[ 2n\pi - \frac{\pi}{2}< w t_0 <  2n\pi + \frac{\pi}{2}\]
We have, from
\[ut_0 - \frac{1}{2} gt_0^2 =0\]
that 
\[t_0 = \frac{2u}{g}\]

  \[ 2n\pi - \frac{\pi}{2}< w \frac{2u}{g} <  2n\pi + \frac{\pi}{2}\]
  Thus, to get heads, $w$ must have the following relation with $u$.
   \[\bigg( 2n-\frac{1}{2}\bigg)\frac{\pi g}{2u}< w  < \bigg( 2n+\frac{1}{2}\bigg)\frac{\pi g}{2u}\]
So, the boundaries separating heads and tails are curves that satisfy 
  \[ w = \bigg( 2n-\frac{1}{2}\bigg)\frac{\pi g}{2u} , n =1,2,3, \hdots\]
  \begin{figure}[H]
\begin{centering}
\includegraphics[width=4in]{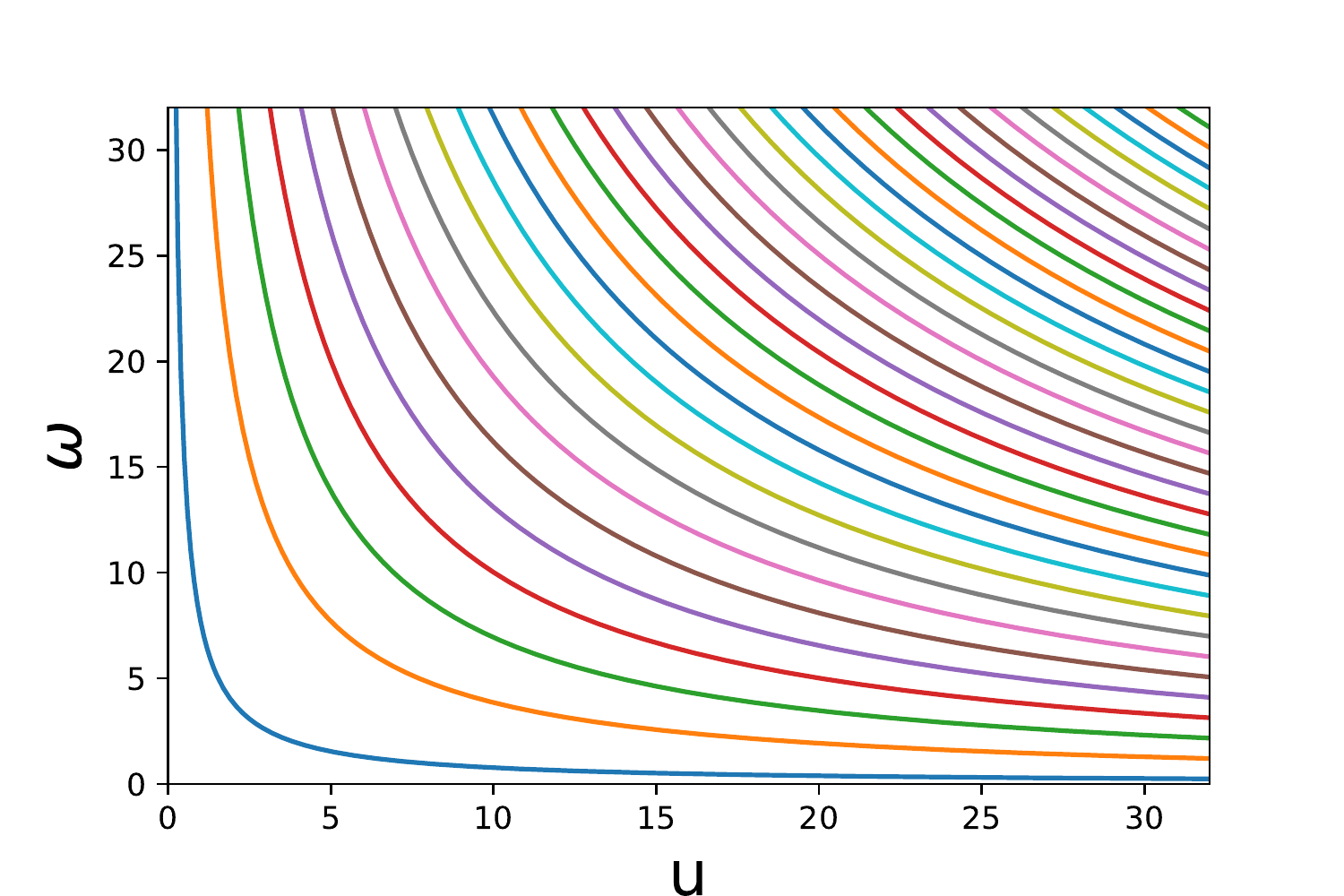}
\caption{Curves that separate heads from tails.}
\end{centering}
\end{figure}
  \section{Keller's Idea}
  
How then is the probability of getting a heads $\frac{1}{2}$?

Let's take a look at the curves in the Fig 1. As $u$ gets large, the curves look like parallel lines that get closer and closer together. In the large $u$ limit,  the lines are so close that it becomes impossible to pick out heads and tails. 
  \begin{figure}[H]
\includegraphics[width=3.5in]{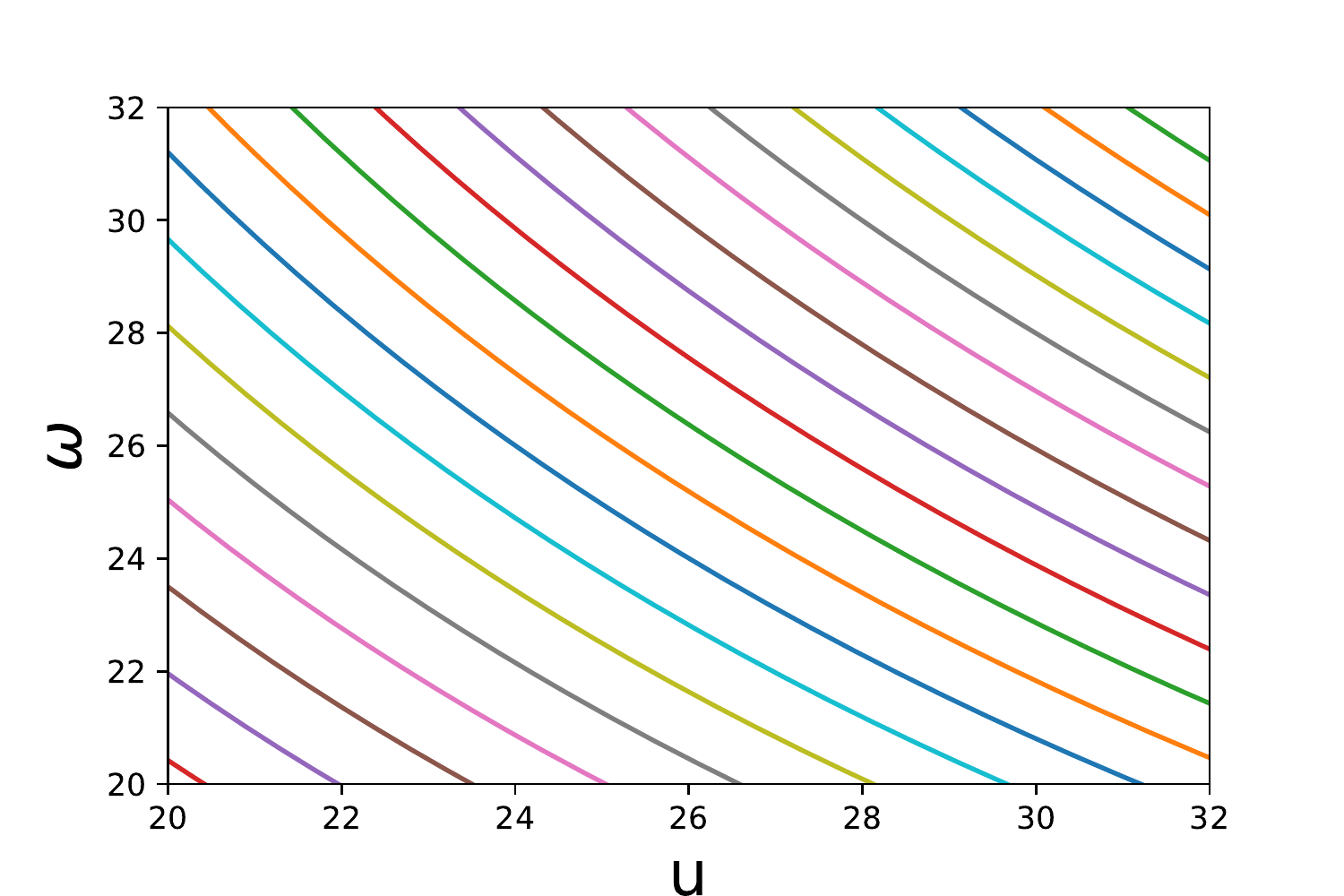}
\caption{The curves for large u look like straight lines. }
\end{figure}
Keller gives a rigorous  and beautiful proof  for why the probability becomes $\frac{1}{2}$ as $u \to \infty$ in the appendix of [1]. In the next section, we explore it with a different ideology.
\section{A lack of 'control'}

If we were able to control our $u$ and $w$ with infinite precision, no matter how close the curves get, we would be able predict exactly if we would get a heads or a tails. Let us aim for an initial velocity of $u_0$ and make an error of magnitude $\Delta u$ in imparting it. Then, the boundary curves become

\begin{align*}
w &= \bigg(2n - \frac{1}{2} \bigg) \frac{\pi g}{2u}\\ 
&= \bigg(2n - \frac{1}{2} \bigg) \frac{\pi g}{2(u_0 + \Delta u)} \\
&\sim  \bigg(2n - \frac{1}{2} \bigg) \frac{\pi g}{2u_0 }(1-\frac{\Delta u }{u_0}) 
\end{align*}

Our approach is to look at the variation of $\omega$ with the change in $u$. Since $u = u_0 + \Delta u$  where $u_0$ is fixed and $\Delta u$ varies, the $\omega$ vs. $u$ can be interpreted as the $\omega$ vs. $\Delta u$ curve and vice versa.

For very large $u_0$,  we have 
\begin{align}
   w \sim  \bigg(2n - \frac{1}{2} \bigg) \frac{\pi g}{2u_0 }\bigg(1-\frac{\Delta u }{u_0}\bigg)  
\end{align}
Then, $\omega$ has a straight line  relation with the error that we make ($\Delta u$). This makes sense because we know that for large $u$, the $w$ vs. $u$ curve is a straight line.

\[w = c+  m\Delta u\]

One would think the slope  of such a line is 
\[\bigg(2n - \frac{1}{2} \bigg) \frac{\pi g}{2u_0^2 } \]
This is not quite right. If n is small compared to $u_0$, this would mean a slope of 0 corresponding to the region around the blue dot in Fig 3.. If n is  very large compared to $u_0^2$, this would mean a slope of infinity corresponding to the red dot. This is why asymptotics have to be done with a careful handle on reality. What we want is the slope near the green dot which corresponds to the straight line approximations. By symmetry, we find that the slope is $-1$.

  \begin{figure}[H]
\includegraphics[width=3.5in]{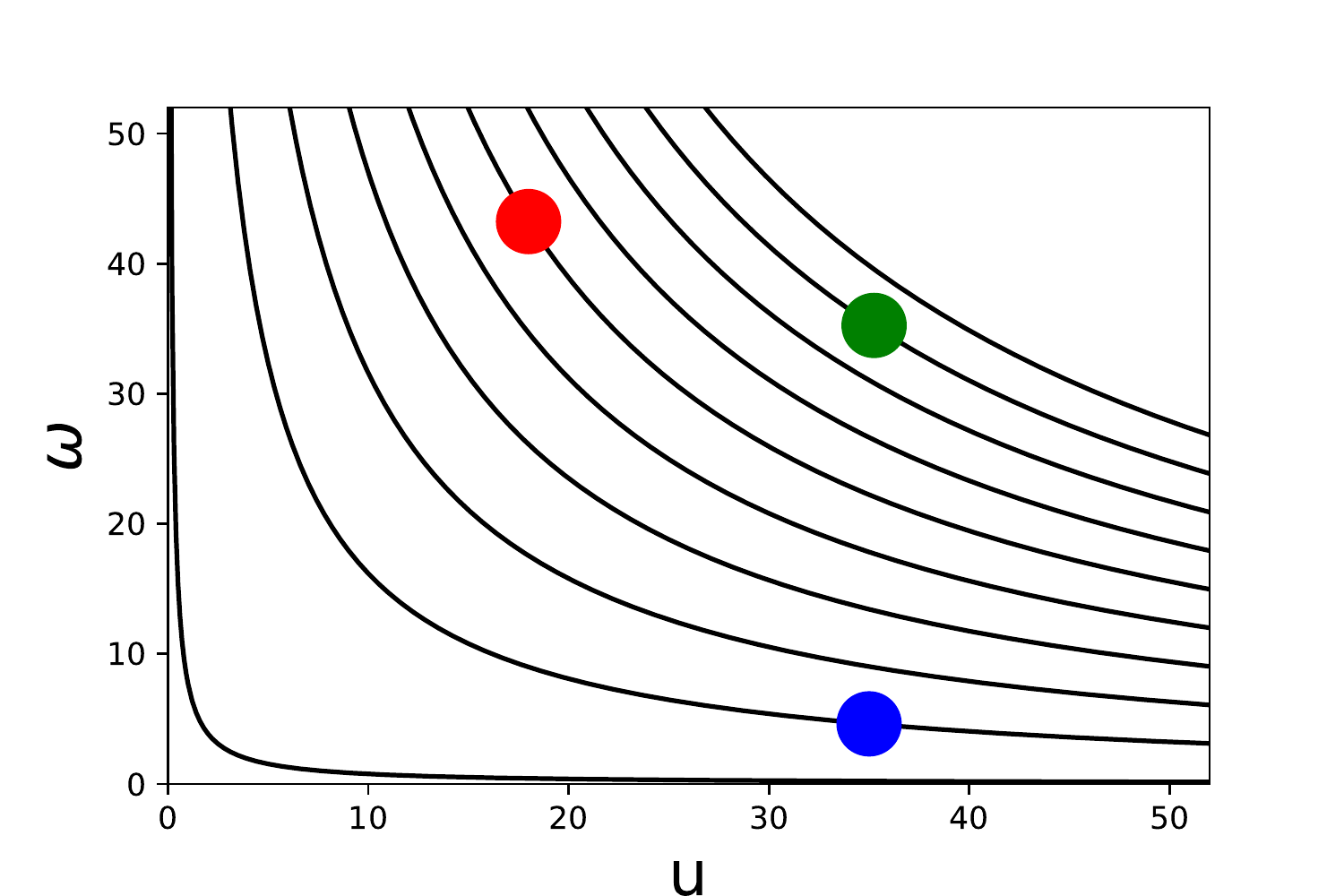}
\caption{Asymptotic approximations need to make sense}
\end{figure}

Finding the slope serves as a nice exercise to remind ourselves where we are working on  but what we really need for the rest of this paper is the offset distance between the successive lines. That distance $a$, shown in Fig 4 and Fig 5. is given by 

\[a = \frac{\pi g}{2 u_0}\]

This allows us to ask the question: What is the maximum permissible error one can make in imparting $u$ and $w$ to the coin? The answer to this is the largest square centered around $u_0$ such that it lies in the region between the two lines.
  \begin{figure}[H]
\includegraphics[width=3.5in]{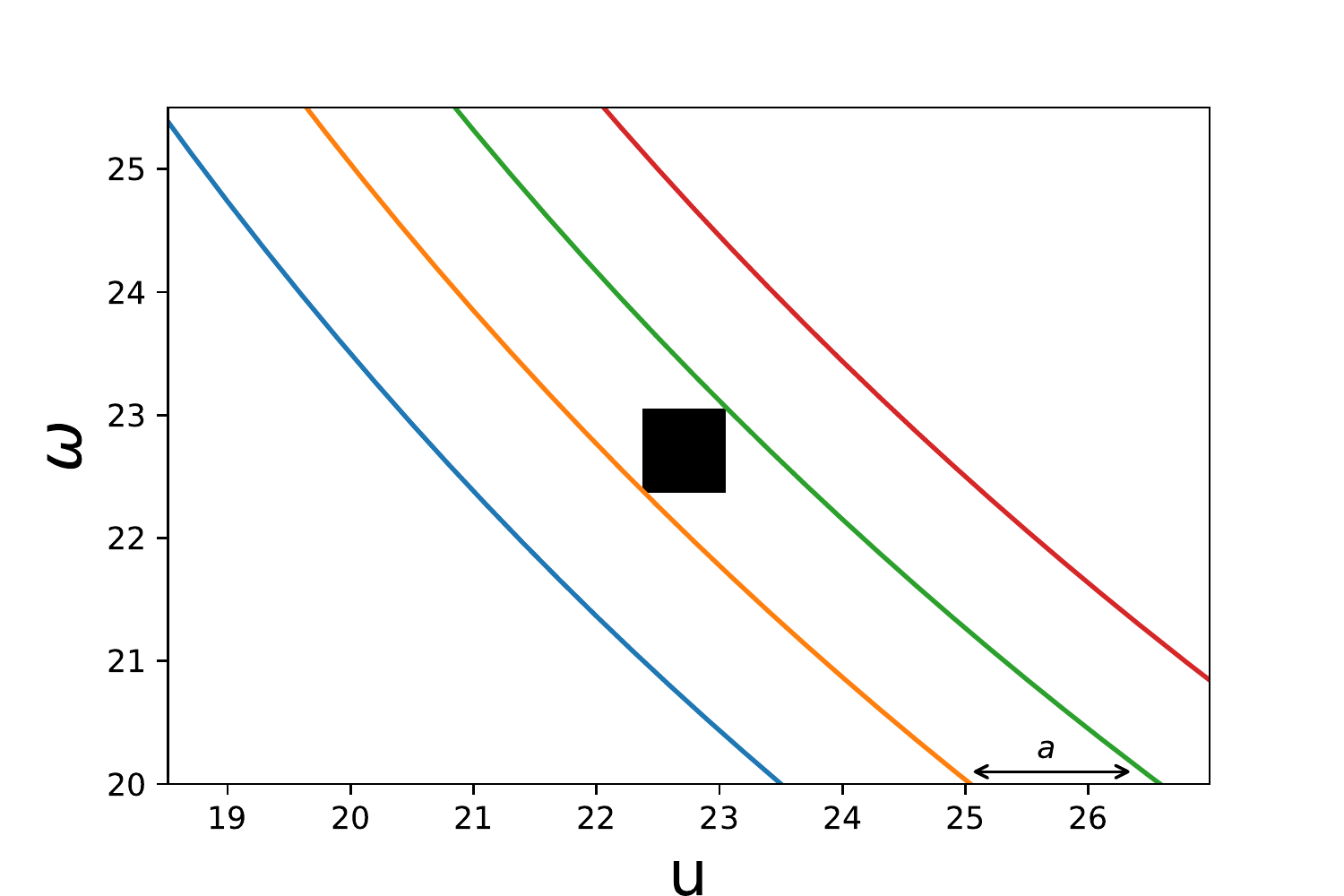}
\caption{Lying inside this black region allows you to get heads with absolute certainty. Note that only 4 curves have been shown here for better clarity.}
\end{figure}
The diagonal of the square is given by
\[d_1 = \frac{a}{\sqrt{2}} = \frac{\pi g}{2 \sqrt{2} u_0}\]
and the sides are
\[\Delta u = \Delta w = \frac{d_1}{\sqrt{2}}= \frac{\pi g}{4u_0}\]
The $\Delta u$ and $\Delta w$ defined above tell the minimum level of control we need to have to make the coin toss deterministic. An inability to control it at this level leads to randomness.

\section{Diagonal definitions}
The diagonal of the square in the previous section was named $d_1$. This is done because we start with some $u_0$, choose the two closest parallel lines on either side of $u_0$ and then find the largest square that fits in between the lines.

  \begin{figure}[H]
\includegraphics[width=3.5in]{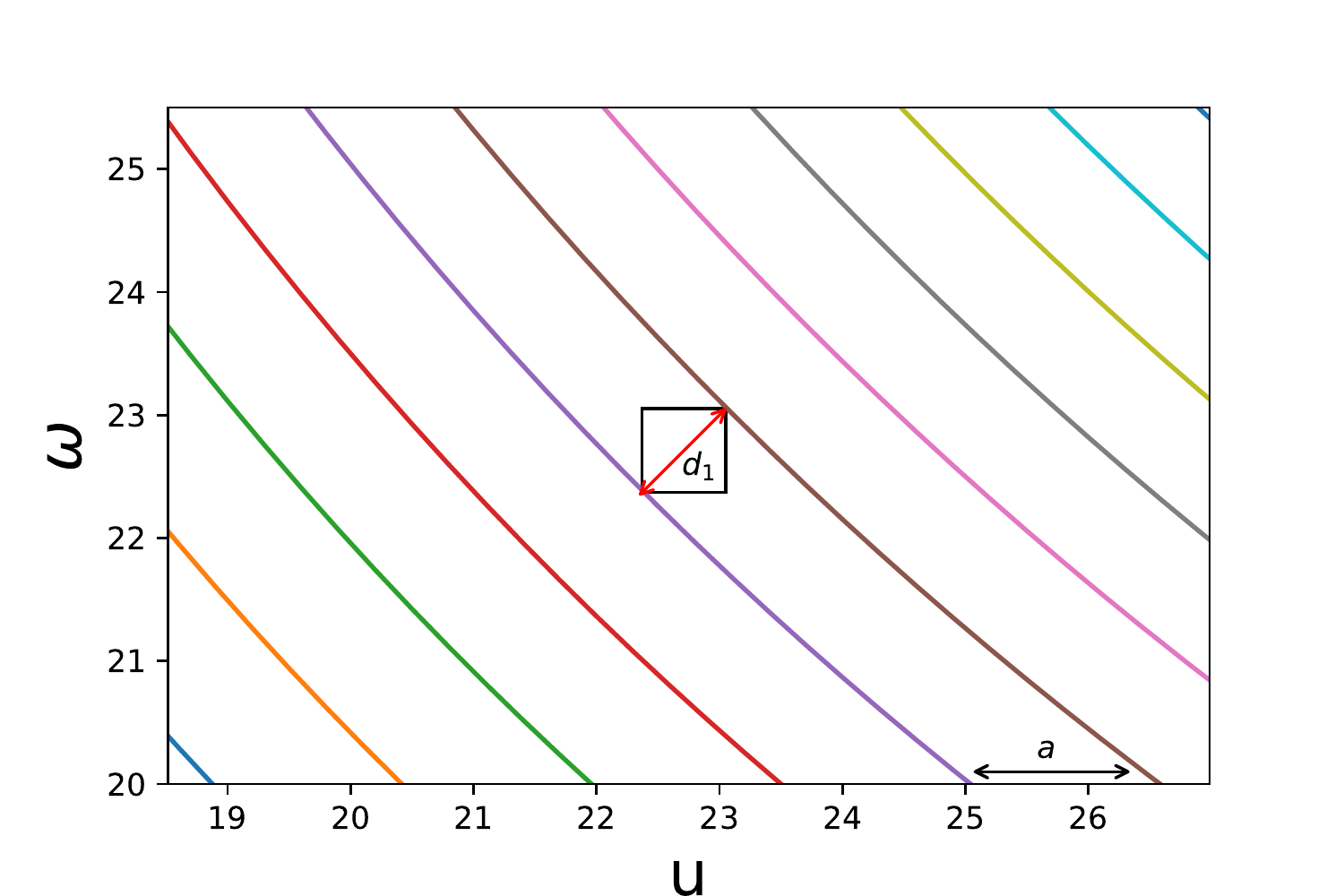}
\caption{$d_1$ is the diagonal of the largest square that lies within the two adjacent lines on either side of a particular $u_0$. }
\end{figure}

We can choose the second closest parallel lines on either side of $u_0$, find the largest square that lies in between them and refer to its diagonal as $d_2$. Similarly, we can find $d_n$ diagonal for larger and larger squares defined this way. $d_1,d_2$ and $d_3$ are highlighted in Fig 6.
  \begin{figure}[H]
\includegraphics[width=3.5in]{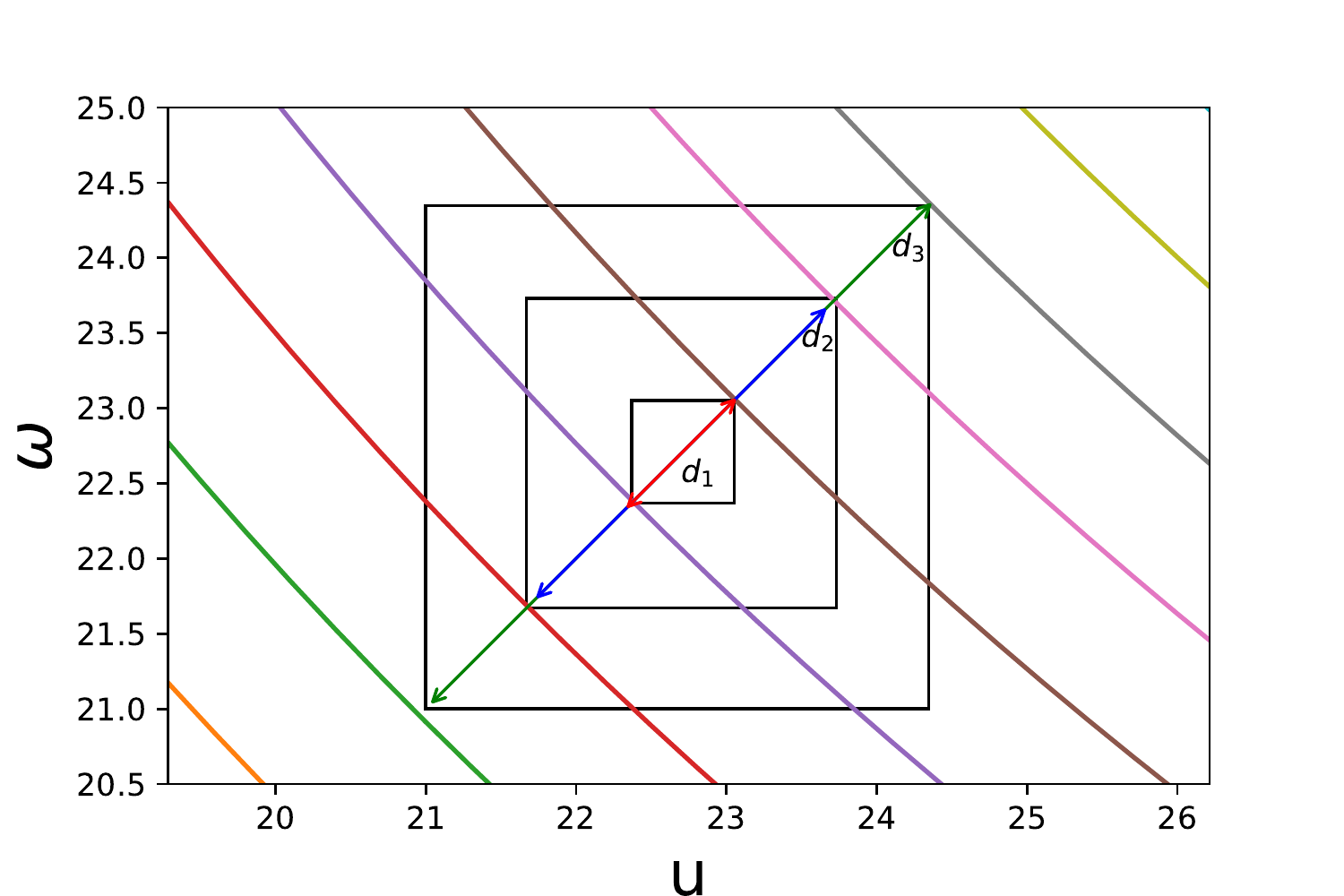}
\caption{$d_2$ and $d_3$ can be found in a similar way to $d_1$. }
\end{figure}
It is not too difficult to show that the $n$th diagonal length is given by
\[d_n = \frac{(2n-1)\pi g}{2 \sqrt{2} u_0} \]

\section{A bigger square}

If we increased the size of square further, the probability of getting heads would decrease from 1.
  \begin{figure}[H]
\includegraphics[width=3.5in]{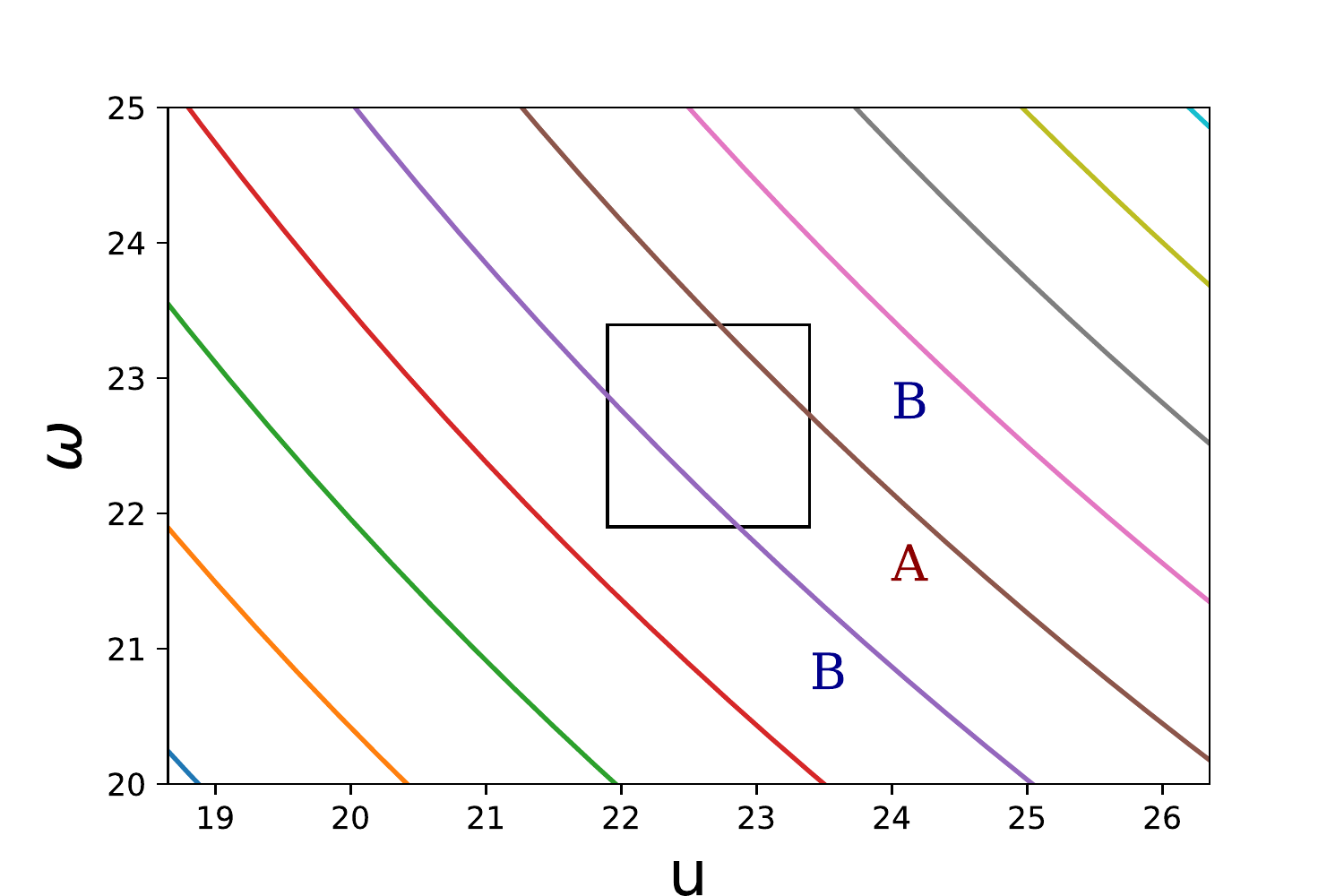}
\caption{When the square of `control' gets outside region A, the coin toss becomes random.  }
\end{figure}
Let's first consider the case when the square is outside region A but still inside region B. So, the probability decreases monotonically with the size of the square. Using the areas of the two regions, we can calculate the probabilities.

Let $d$ be the diagonal of the square. Clearly, $d_1<d<d_2$.
The total area of the square is \[ \frac{1}{2}d^2\]
Also, the total area of each of the two triangles in region B is  \[ \frac{1}{4}(d-d_1)^2\].
So, the combined area of the two triangles is \[\frac{1}{2}(d-d_1)^2\]
Hence, the probability of tails is given by the ratio of the areas.
\[P_T = \frac{d-d_1^2}{d^2}\]
and that of heads is given by
\[P_H = 1-  \frac{d-d_1^2}{d^2}\]

  \begin{figure}[H]
\includegraphics[width=3.5in]{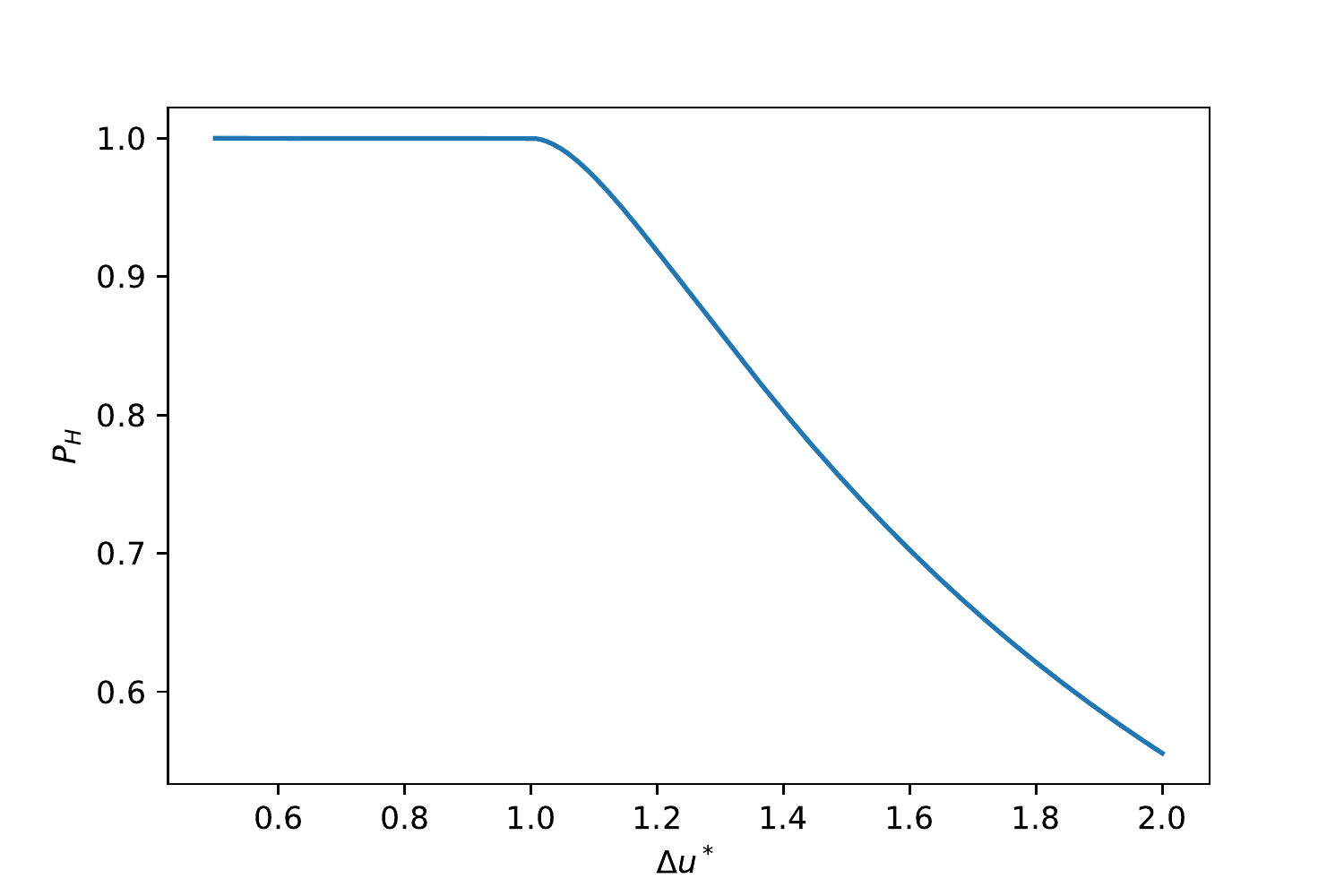}
\caption{Probability of heads as the square gets into region B. }
\end{figure}
In general, we wish to understand the probability distribution for situations like in the Fig 7 below.
  \begin{figure}[H]
\includegraphics[width=3.5in]{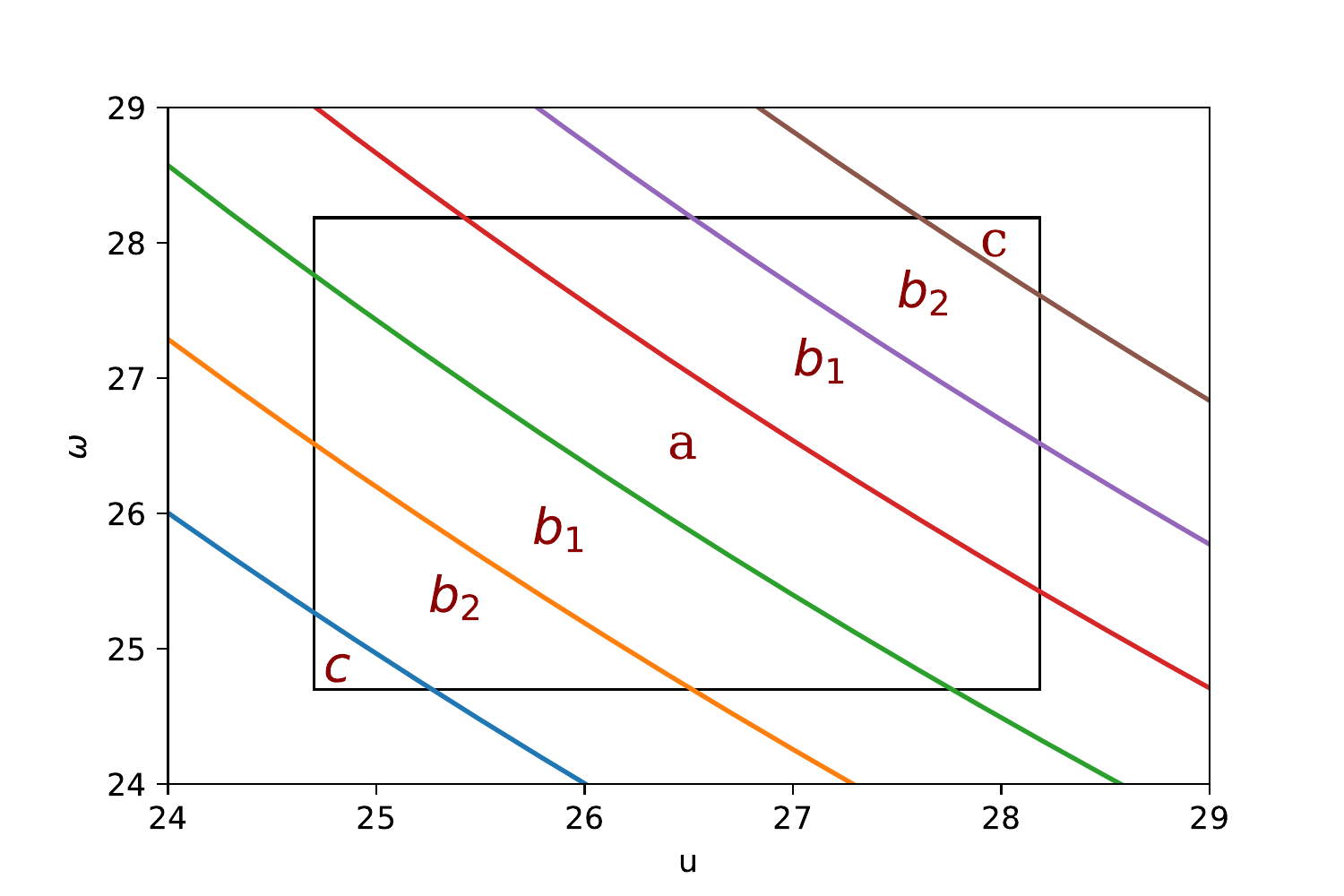}
\caption{A more complicated square.  }
\end{figure}

\[\text{Area of region a } = \frac{1}{2}d^2 - \frac{1}{2}(d-d_1)^2\]
\[\text{Area of region $b_n$ } = \frac{2d-d_{n-1}-d_n}{2} d_1 \]
\[\text{Area of region c } =  \frac{1}{4}(d-d_N)^2\]

We can use these areas to find the probability of getting a heads as the size of the square increases. It is shown in Fig 7.

  \begin{figure}[H]
\includegraphics[width=3.5in]{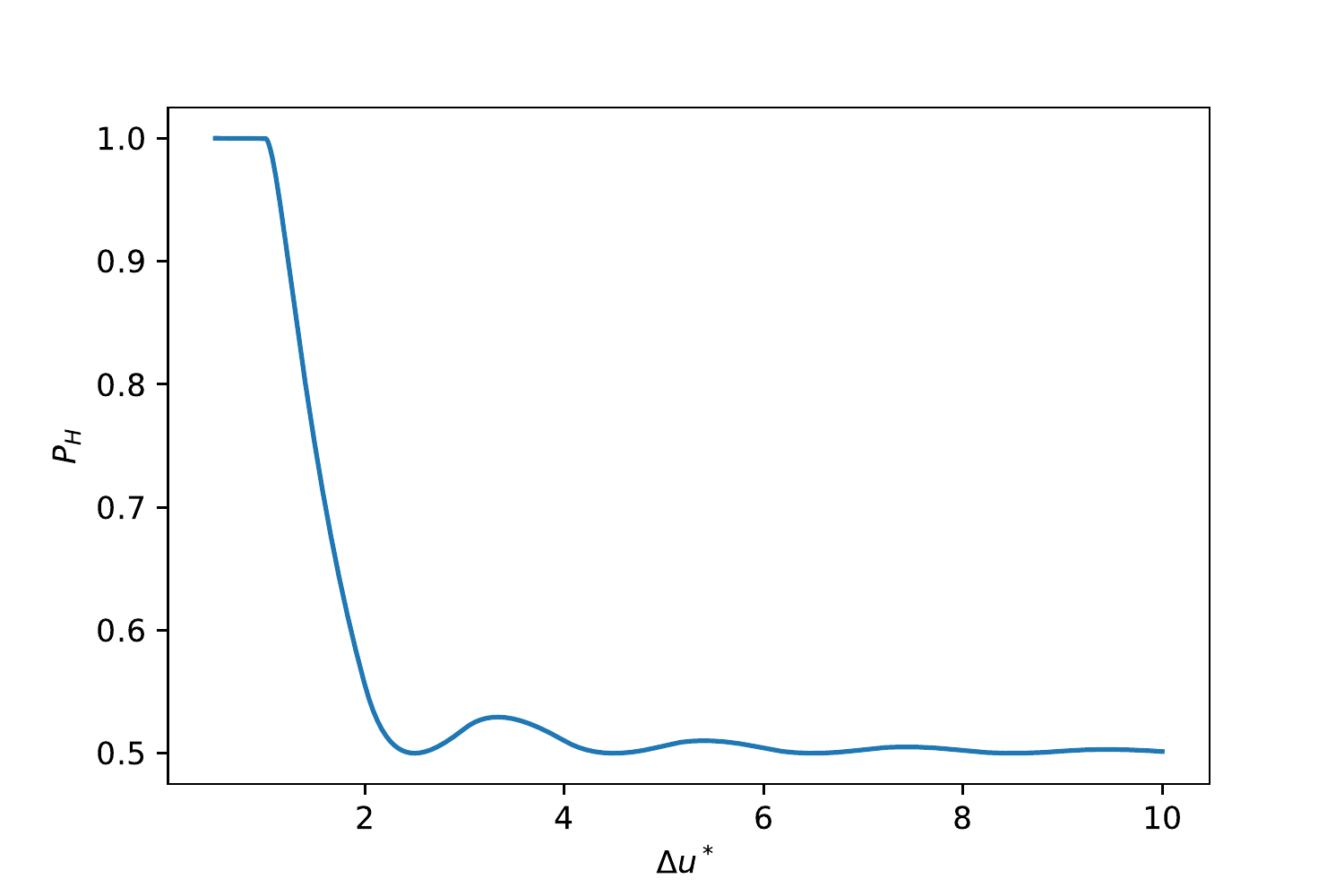}
\caption{Probability of heads as the error (normalized) in u increases.}
\end{figure}

Fig 8 allows us to see how as you lose control over the impulse imparted to the coin, the probability of heads eventually goes to 0.5. Interestingly, the probability never gets below 0.5. This means that you are ever slightly more likely to get the side you started with when you tossed the coin. Persi Diaconis found a similar bias in the toss of a usual coin in [3]. The probability of heads in that study was 0.51! (Not a factorial, an exclamation).

\section{Real life scenarios}

From personal experimentation, a coin is tossed with a velocity near 4m/s which does not qualify it to be in the $u \to \infty$ region. Furthermore, measuring $\omega$ is extremely difficult as expressed by Diaconis in his Numberphile videos.

\section{Conclusions}

This article tries to explore how the meaning of what we call random or deterministic is related to how much control we have over it. It also hopefully leaves readers with the impression that they can never look at the simple coin toss the same way again. Something we always thought was random could be deterministic based on our level of control. This insight also serves as a cure for one of the dilemmas faced by people introduced to statistical mechanics. The motion of particles is described by Newton's laws and hence should be completely deterministic. That motion however takes place at such a small scale and with such a large number of interactions that our lack of control or `precision'  makes it a random process.

\end{document}